\documentclass[10pt,a4paper,english]{article}\usepackage[]{graphicx}\usepackage[]{xcolor}
\makeatletter
\def\maxwidth{ %
  \ifdim\Gin@nat@width>\linewidth
    \linewidth
  \else
    \Gin@nat@width
  \fi
}
\makeatother

\definecolor{fgcolor}{rgb}{0.345, 0.345, 0.345}

\usepackage{framed}
\makeatletter
 {\par\unskip\endMakeFramed%
 \at@end@of@kframe}
\makeatother

\definecolor{shadecolor}{rgb}{.97, .97, .97}
\definecolor{messagecolor}{rgb}{0, 0, 0}
\definecolor{warningcolor}{rgb}{1, 0, 1}
\definecolor{errorcolor}{rgb}{1, 0, 0}

\usepackage{alltt}

\usepackage[hyphens]{url}
\usepackage{amsmath}
\usepackage{amsfonts}
\usepackage{amssymb}
\usepackage{amsthm}

\theoremstyle{plain}

\theoremstyle{definition}

\theoremstyle{remark}

\PassOptionsToPackage{hyphens}{url}\usepackage{hyperref}
\usepackage{hyperref}
\usepackage[square,numbers]{natbib}
\usepackage{orcidlink}

\usepackage[newitem,newenum,increaseonly]{paralist}

\usepackage[left=2.00cm,right=1.00cm]{geometry}
\usepackage{tikz}
\usetikzlibrary{angles, quotes}

\usepackage[notheorems]{SharpeR}

\usepackage{smm}

    \usepackage{color} 
    \usepackage{enumerate} 
    \usepackage{amsmath} 
    \usepackage{amssymb} 
		\usepackage{fancyvrb} 
    \usepackage{hyperref}

    \definecolor{orange}{cmyk}{0,0.4,0.8,0.2}
    \definecolor{darkorange}{rgb}{.71,0.21,0.01}
    \definecolor{darkgreen}{rgb}{.12,.54,.11}
    \definecolor{myteal}{rgb}{.26, .44, .56}
    \definecolor{gray}{gray}{0.45}
    \definecolor{lightgray}{gray}{.95}
    \definecolor{mediumgray}{gray}{.8}
    \definecolor{inputbackground}{rgb}{.95, .95, .85}
    \definecolor{outputbackground}{rgb}{.95, .95, .95}
    \definecolor{traceback}{rgb}{1, .95, .95}
    \definecolor{red}{rgb}{.6,0,0}
    \definecolor{green}{rgb}{0,.65,0}
    \definecolor{brown}{rgb}{0.6,0.6,0}
    \definecolor{blue}{rgb}{0,.145,.698}
    \definecolor{purple}{rgb}{.698,.145,.698}
    \definecolor{cyan}{rgb}{0,.698,.698}
    \definecolor{lightgray}{gray}{0.5}
    
    \definecolor{darkgray}{gray}{0.25}
    \definecolor{lightred}{rgb}{1.0,0.39,0.28}
    \definecolor{lightgreen}{rgb}{0.48,0.99,0.0}
    \definecolor{lightblue}{rgb}{0.53,0.81,0.92}
    \definecolor{lightpurple}{rgb}{0.87,0.63,0.87}
    \definecolor{lightcyan}{rgb}{0.5,1.0,0.83}

    \definecolor{incolor}{rgb}{0.0, 0.0, 0.5}
    \definecolor{outcolor}{rgb}{0.545, 0.0, 0.0}

    \hypersetup{
      breaklinks=true,  
      colorlinks=true,
      urlcolor=blue,
      linkcolor=darkorange,
      citecolor=darkgreen,
      }
\IfFileExists{upquote.sty}{\usepackage{upquote}}{}
\begin{document}

\title{The Sherman-Morrison-Markowitz Portfolio}
\author{\orcidlink{0000-0002-4197-6195} Steven E. Pav \thanks{\email{steven@gilgamath.com}}}

\maketitle

\begin{abstract}
We show that the Markowitz portfolio is a scalar multiple of another portfolio which replaces the covariance with the second moment matrix,
via simple application of the Sherman-Morrison identity.
Moreover it is shown that when using conditional estimates of the first two moments, this ``Sherman-Morrison-Markowitz'' portfolio 
solves the standard unconditional portfolio optimization problems.
We argue that in multi-period portfolio optimization problems it is more natural to replace variance and covariance with
their uncentered counterparts.
We extend the theory to deal with constraints in expectation, where we find a decomposition of squared effects into 
spanned and orthogonal components.
Compared to the \txtMP, the \txtSMMP down-levers by a small amount that depends on the conditional \txtSRSQOPT; 
the practical impact will be fairly small, however.
We present some example use cases for the theory.

\end{abstract}

\section{Introduction}

The \txtMP plays a central role in the theory of quantitative portfolio management, as it solves several varieties of portfolio
optimization problems.
\cite{markowitz1952portfolio,Ruppert:2006}
Yet despite the theoretical backing, the \txtMP has limited practical reach, largely due to estimation error.
\cite{michaud1989markowitz,demiguel2009optimal,pav2014qbounds}
Real world portfolio optimization also usually imposes constraints which do not admit closed form solutions,
leading to a gap in the theoretical understanding of practical portfolio construction methodology.

The \txtMP solves both the Sharpe maximization and mean-variance optimization problems in the single period model of investment.
In this note we consider the multi-period (or ``intertemporal'') variants of these problems,
where one has access to some features upon which one can condition the investment decision.
The objective of the multi-period investment problem is to construct some function from the features to portfolios which optimizes
the unconditional multi-period moments of returns.
The main results of this note are:
\begin{compactenum}
\item The \txtMP conditional on the features is not optimal for the multi-period problem, rather a closely related portfolio is.
\item The two portfolios are scaled versions of each other, and are connected by the Sherman-Morrison formula for the inverse of a
  rank-one update of a matrix.
\item Under the multi-period problem it is more natural to work with the second moment and second moment matrix than the variance
  and covariance matrix. 
  Moreover, it is more natural to work with the (squared) \txtHR, defined by \citeauthor{ern2020hansen} as mean divided by square root of second moment, 
  than the \txtSR. \cite{ern2020hansen}
\end{compactenum}
Moreover, because of the rescaling property, our ``\txtSMMP'' is optimal in the single period case as
well, even though the \txtMP is not optimal in the multi-period case.
We seek to turn portfolio theory not quite upside down, but perhaps sideways, by suggesting that the covariance and variance in 
\txtMP and \txtSR be replaced by their uncentered variants.

The Sherman-Morrison formula relates the inverse of a rank-one update of a matrix to the original matrix inverse.  \cite{hager_updating_1989}
Letting \pvmu and \pvsig be the expected value and covariance matrix of returns,
this formula implies that 
\begin{equation}
\minvParens{\pvsig + \pvmu\trpvmu}\pvmu = \frac{1}{1 + \trpvmu\pvinvsig\pvmu} \pvinvsig\pvmu,
\label{eqn:smm_relation}
\end{equation}
which can also be verified by direct multiplication of both sides by the second moment matrix ${\pvsig + \pvmu\trpvmu}$.
The vector $\pvinvsig\pvmu$ is the \txtMP, which is usually scaled by some positive constant depending on constraints or
other particulars of the problem at hand, while $\minvParens{\pvsig + \pvmu\trpvmu}\pvmu$ is the \txtSMMP.
The optimality of the latter comes not from chosing a different portfolio direction for each period, but
rather by downscaling by $\wrapParens{1 + \trpvmu\pvinvsig\pvmu}^{-1}$ in each period, compared to the \txtMP.

In this note we make a number of departures from the typical assumptions of portfolio problems.
For one we discard the assumption of a fully invested portfolio, the so-called ``self-financing condition''; 
one can view our problem as an optimization over risky assets while the remainder of one's wealth, long or short, 
is implicitly invested in the risk-free asset.  

Secondly we focus on maximizing the unconditional, or multi-period, \txtSR, or the multi-period mean-variance objective.
This is somewhat unorthodox among academic papers, but should not be unnatural for practitioners.
Indeed, common practice is to backtest trading systems which make several trading decisions, 
then estimate the \txtSR from the entire backtest period.
Investors might ask how such a trading system would perform in some economic crisis in the distant past, 
implicitly performing some kind of mental averaging over an entire economic cycle.
Moreover, investors often surrender their capital to fund managers for longer periods than horizon of a a single investment decision,
so managers should be cognizant of the multi-period objective.

Thirdly the constraints that we impose--an overall cap on unconditional risk, hedging constraints--are couched in terms of
long term expectations, rather than single-period constraints. 
These choices were expedient, as they make the math work out neatly, but may be unsatisfactory since they can violate
risk constraints in the single period problem. 
The conditional constrants can be achieved by assuming a set of basis portfolio functions.
However, our exposition can only handle the case where that set is finite; 
further work is required for the case of an infinite set of basis functions.


\section{Unconditional Sharpe Maximization}

Let $\vreti[t]$ be a \nasst-length vector of the (percent) returns of some assets.
Suppose that prior to the investment decision you observe a random $\nftr$-vector of ``features,'' $\ftr[t]$ which can guide your investment decision.
Let the density of $\ftr[t]$ be $\fdens{\ftr[t]}$.
Express the conditional mean and second moment of $\vreti[t]$, conditional on $\ftr[t]$, as the following functions:
\begin{align}
  \pvmufnc{\ftr[t]} & \defeq \Econd{\vreti[t]}{\ftr[t]},\\
  \pvsmfnc{\ftr[t]} & \defeq \Econd{\vreti[t]\trvreti[t]}{\ftr[t]}.
\end{align}
Based on these we define the conditional covariance and optimal \txtSR as the functions
\begin{align}
\pvsigfnc{\ftr[t]} & \defeq \pvsmfnc{\ftr[t]} - \pvmufnc{\ftr[t]}\trpvmufnc{\ftr[t]},\\
  \psnroptfnc{\ftr[t]} & \defeq \sqrt{\trpvmufnc{\ftr[t]}\pvinvsigfnc{\ftr[t]}\pvmufnc{\ftr[t]}}.
\end{align}

Suppose that in response to observing $\ftr[t]$, you allocate $\pportwfnc{\ftr[t]}$ of your wealth into each of the \nasst assets,
for $\pportwfnc{\cdot}$ selected from some set of acceptable functions which takes $\reals{\nftr}$ to $\reals{\nasst}$.
The expected return of your portfolio conditional on observing $\ftr[t]$ is then $\trpportwfnc{\ftr[t]}\pvmufnc{\ftr[t]}$, 
and the conditional expected squared return of your portfolio is $\trpportwfnc{\ftr[t]}\pvsmfnc{\ftr[t]}\pportwfnc{\ftr[t]}$.
The \emph{unconditional} mean return, second moment of return, variance of return and risk
are 
defined as
\begin{align}
  \pmufnc{\pportw} & \defeq \int \trpvmufnc{\ftr}\pportwfnc{\ftr}\fdens{\ftr}\dx[\ftr],\\
  \psmfnc{\pportw} & \defeq \int \trpportwfnc{\ftr}\pvsmfnc{\ftr}\pportwfnc{\ftr}\fdens{\ftr}\dx[\ftr],\\
  \pvarfnc{\pportw} & \defeq \psmfnc{\pportw} - \wrapParens{\pmufnc{\pportw}}^2,\\
  \priskfnc{\pportw} & \defeq \sqrt{\pvarfnc{\pportw}}.
\end{align}

Note that these functionals are homogeneous of certain degree.
That is for scalar $c$ define $c \scalop \pportw$ as the function $\pportw$ scaled by $c$:
$$
\funcit{\wrapParens{c\scalop\pportw}}{\ftr[t]} = c \funcit{\pportw}{\ftr[t]}.
$$
We can think of $c \scalop \pportw$ as the allocation $\pportw$ scaled up (or down) by $c$.
Then for scalar $c$ we have
\begin{align*}
\pmufnc{c\scalop\pportw} &= c \pmufnc{\pportw},&
\psmfnc{c\scalop\pportw} &= c^2 \psmfnc{\pportw},\\
\pvarfnc{c\scalop\pportw} &= c^2 \pvarfnc{\pportw},&
\priskfnc{c\scalop\pportw} &= \abs{c}\priskfnc{\pportw}.
\end{align*}

We are interested portfolio functions $\pportwfnc{\ftr[t]}$ which maximize the \txtSR
$\wrapParens{\pmufnc{\pportw} - \rfr}/{\priskfnc{\pportw}},$
for some risk-free rate $\rfr$, or which maximize the mean-variance objective
\begin{equation}
\nonumber
  \pmvofnc{\pportw} \defeq \pmufnc{\pportw} - \half[\riskt]\pvarfnc{\pportw},
\end{equation}
for some risk intolerance $\riskt > 0$.
We define the \txtSR functional for zero risk-free rate as
\begin{equation}
\nonumber
  \psnrfnc{\pportw} \defeq \frac{\pmufnc{\pportw}}{\priskfnc{\pportw}},
\end{equation}
which is positively homogeneous of degree zero.
That is, $\psnrfnc{c\scalop\pportw}=\sign{c}\psnrfnc{\pportw}$.
The mean variance objective $\pmvofnc{\pportw}$ is not homogeneous,
which indicates that \riskt is not unitless, and may depend on the investment horizon.

We consider portfolio functions which are members of \Ltwo, which is defined as the set of of 
\pportwfnc{\ftr} such that
$$
\int \trpportwfnc{\ftr}\pportwfnc{\ftr}\fdens{\ftr}\dx[\ftr] < \infty
$$
Moreover, we will assume that $\pportw$ are drawn from some positive cone $\cone \subseteq\Ltwo$.
A positive cone is a set of functions such that if $\pportw \in \cone$ then 
$c\scalop\pportw \in \cone$ for every positive constant $c$.
In the most generic setting \cone will be \Ltwo itself, but we consider portfolio constraints in the sequel
where \cone is some set of acceptable allocations.

Now for $\rfr \ge 0$ and $\Rbuj > 0$ consider the \txtSR optimization problem with a risk budget:
\begin{equation}
\max_{\pportw\in\cone \,:\, \priskfnc{\pportw} \le \Rbuj} \frac{\pmufnc{\pportw} - \rfr}{\priskfnc{\pportw}}.
  \label{problem:opt_portsr}
\end{equation}
Here \rfr is the risk-free rate, and \Rbuj is the maximum allowable risk.
By a series of transformations we aim to show that the solution to this optimization problem
is related to the solution of the following problem:
\begin{equation}
  \max_{\pportw\in\cone \,:\, \psmfnc{\pportw} = 1} \pmufnc{\pportw}.
  \label{problem:opt_portsr_final}
\end{equation}
By ``related to'' we mean that one can rescale a solution to \problemref{opt_portsr_final} in a formulaic
way by a positive scalar to arrive at a solution to \problemref{opt_portsr}, thus one can instead focus on the latter problem.

First we show that any solution to \problemref{opt_portsr} will saturate the risk budget if $\rfr > 0$.
If $\rfr = 0$, then a solution to \problemref{opt_portsr} can be positively rescaled to saturate the risk
budget without changing the objective, and thus we can replace that problem with the equivalent problem,
\begin{equation}
\max_{\pportw\in\cone \,:\, \priskfnc{\pportw} = \Rbuj} \frac{\pmufnc{\pportw} - \rfr}{\priskfnc{\pportw}}.
  \label{problem:opt_portsr_II}
\end{equation}

Towards a contradiction, let \pportwopt optimize \problemref{opt_portsr}, but suppose that
$\priskfnc{\pportwopt} < \Rbuj$.
Then there is scalar $c = 1 + \lambda$ with $\lambda > 0$ such that 
$\priskfnc{\wrapParens{1+\lambda}\scalop\pportwopt} = \Rbuj$. 
But then, using homogeneity
\begin{align*}
\frac{\pmufnc{\wrapParens{1+\lambda}\scalop\pportwopt} - \rfr}{\priskfnc{\wrapParens{1+\lambda}\scalop\pportwopt}} 
&= \frac{\wrapParens{1+\lambda}\pmufnc{\pportwopt} - \rfr}{\wrapParens{1+\lambda}\priskfnc{\pportwopt}}\\
&= \frac{\wrapParens{1+\lambda}\pmufnc{\pportwopt} - \wrapParens{1+\lambda}\rfr + \lambda\rfr}{\wrapParens{1+\lambda}\priskfnc{\pportwopt}}\\
&= \frac{\pmufnc{\pportwopt} - \rfr}{\priskfnc{\pportwopt}} + \frac{\lambda\rfr}{\wrapParens{1+\lambda}\priskfnc{\pportwopt}}.
\end{align*}
Now this is at least equal to the objective of $\pportwopt$ and strictly greater than it if $\rfr > 0$.
This would be a contradiction to the optimality of $\pportwopt$ if $\rfr > 0$, and otherwise establishes that we can,
without loss of generality when $\rfr = 0$, solve \problemref{opt_portsr_II} instead.

But with the risk budget saturated, trivially we can rewrite that optimization problem as 
\begin{equation}
\max_{\pportw\in\cone \,:\, \priskfnc{\pportw} = \Rbuj} \frac{\pmufnc{\pportw} - \rfr}{\Rbuj}.
\label{problem:opt_portsr_III}
\end{equation}
Given that $\Rbuj$ is fixed and positive, it suffices to instead solve the problem
\begin{equation}
\max_{\pportw\in\cone \,:\, \priskfnc{\pportw} = \Rbuj} \pmufnc{\pportw}.
\label{problem:opt_portsr_IV}
\end{equation}
Because this optimization saturates the risk budget, we can shift it to the denominator of the objective
to arrive at the equivalent problem
\begin{equation}
  \max_{\pportw\in\cone \,:\, \priskfnc{\pportw} = \Rbuj} \frac{\pmufnc{\pportw}}{\priskfnc{\pportw}}.
\label{problem:opt_portsr_V}
\end{equation}

Now consider the following diagram, where $c$ is some positive scalar:

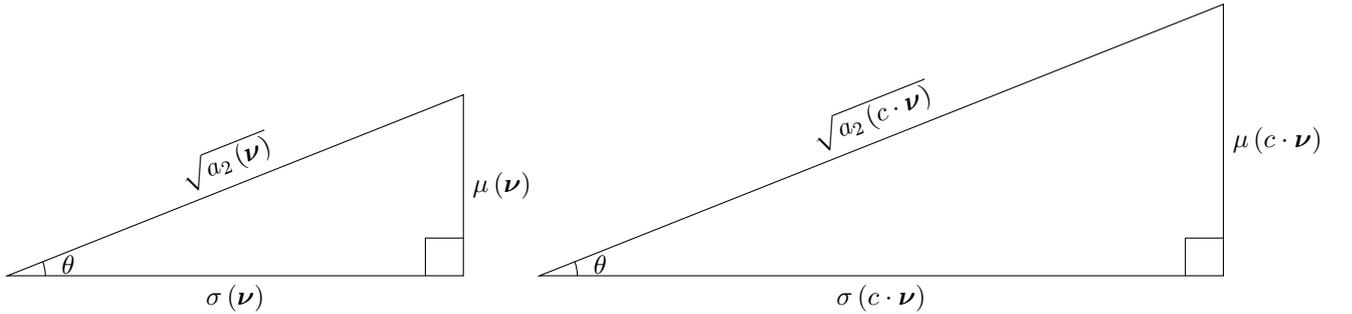
\begin{figure}[h]
  \centering
\begin{tikzpicture}
  \draw
  (1,0) coordinate (O) 
  -- (7,0) coordinate (A) 
  node[midway, below]{$\priskfnc{\pportw}$}
  -- (7,2.4) coordinate (B) 
  node[midway, right]{$\pmufnc{\pportw}$}
  -- cycle
  node[midway, above, sloped]{$\sqrt{\psmfnc{\pportw}}$}
  pic[draw, black] {right angle=O--A--B}
  (1.8,0.15) coordinate (T) node[]{$\theta$}
  pic[draw, black] {angle=A--O--B}

  (8,0) coordinate (OO) 
  -- (17,0) coordinate (AA) 
  node[midway, below]{$\priskfnc{c\scalop\pportw}$}
  -- (17,3.6) coordinate (BB) 
  node[midway, right]{$\pmufnc{c\scalop\pportw}$}
  -- cycle
  node[midway, above, sloped]{$\sqrt{\psmfnc{c\scalop\pportw}}$}
  pic[draw, black] {right angle=OO--AA--BB}
  (8.8,0.15) coordinate (TT) node[]{$\theta$}
  pic[draw, black] {angle=AA--OO--BB}
  ;
\end{tikzpicture}
\end{figure}

\problemref{opt_portsr_V} requires us to find the $\pportw$ that maximizes the tangent of $\theta$,
equivalently that maximizes $\theta$, subject to an equality constraint on the horizontal leg.
However, it is easily seen that to solve that problem it suffices to maximize $\theta$ subject
to any other equality constraint that is positively homogeneous of degree 1, then rescale the result. 
That is, to solve \problemref{opt_portsr_V} it suffices to instead solve
\begin{equation}
  \max_{\pportw\in\cone \,:\, \psmfnc{\pportw} = 1} \pmufnc{\pportw},
\label{problem:opt_portsr_VI}
\end{equation}
and rescale the optimal portfolio function to achieve the requisite equality constraint.

We sketch the proof for any skeptics:
let $\pportwopt$ be a solution to \problemref{opt_portsr_VI}, then consider scalar $c$ such that 
$$
\priskfnc{c\scalop\pportwopt} = \Rbuj.
$$
Now suppose $\pportw[1]$ is some other portfolio such that 
$\priskfnc{\pportw[1]} = \Rbuj$, but which gives strictly larger expected return:
$$
\pmufnc{\pportw[1]} > \pmufnc{c\scalop\pportwopt}.
$$
Now
\begin{align*}
\pvarfnc{c\scalop\pportwopt} &= \pvarfnc{\pportw[1]},\\
\psmfnc{c\scalop\pportwopt} - \wrapParens{\pmufnc{c\scalop\pportwopt}}^2 &= 
\psmfnc{\pportw[1]} - \wrapParens{\pmufnc{\pportw[1]}}^2,\\
c^2 = \psmfnc{c\scalop\pportwopt} &> \psmfnc{\pportw[1]},
\end{align*}
where the last inequality follows from the inequality on means. 
Thus there is some $k > c^{-1}$ such that $\psmfnc{k \pportw[1]} = 1$. 
But then $\pmufnc{k\pportw[1]} = k \pmufnc{\pportw[1]} > c^{-1} \pmufnc{c\scalop\pportwopt} = \pmufnc{\pportwopt},$
which is a contradiction to $\pportwopt$ solving \problemref{opt_portsr_VI}.
Thus $c\scalop\pportwopt$ must be a solution to \problemref{opt_portsr_V}.

The reverse implication follows similarly, establishing an equivalence between \problemref{opt_portsr_VI} and
\problemref{opt_portsr_V}, up to scaling by some constant. 
Moreover, this shows that to solve \problemref{opt_portsr}, it suffices to solve \problemref{opt_portsr_VI} and rescale
the portfolio to saturate the risk constraint.

\subsection{Unconstrained Case}

We now consider solutions to \problemref{opt_portsr_VI} when \cone is the entire space of square-integrable functions.
Without the cone constraint, this 
is an example of an \emph{isoperimetric problem}.  \cite{maccluer2005calculus}
That is, we can write it as
\begin{equation}
\nonumber
  \max_{\pportwfnc{\ftr}} \int \funcit{L}{\ftr, \pportw} \dx[\ftr] \quad\mbox{s.t.}\quad
  \int \funcit{M}{\ftr, \pportw} \dx[\ftr] = 1,
\end{equation}
for certain functions $L$ and $M$.
To be specific
\begin{align*}
  \funcit{L}{\ftr, \pportw} &= \tr{\pportw}\pvmufnc{\ftr}\fdens{\ftr},\\
  \funcit{M}{\ftr, \pportw} &= \tr{\pportw}\pvsmfnc{\ftr}\pportw \fdens{\ftr}.
\end{align*}
These functions have no dependence on the gradient of $\pportw$, which makes them atypical of problems in the calculus of
variations,
and the solution to our problem is relatively simple.
A necessary condition for a solution $\pportwfnc{\ftr[t]}$ is that either $M$ satisfies the Euler-Lagrange equation,
or for some $\lambda$ the linear combination $N = L + \lambda M$ satisfies the Euler-Lagrange equation.
Without any dependence on the gradient of $\pportw$ the Euler-Lagrange equation reduces to what looks like a typical elementary
calculus necessary condition for an optimum, namely at a solution either
\begin{equation}
\gradof[\pportw]{M} = \vzero, \quad\mbox{or}\quad \gradof[\pportw]{N} = \vzero.
\end{equation}
Given the definition of the objective and constraint, these reduce to 
\begin{equation}
  2 \pvsmfnc{\ftr}\pportwfnc{\ftr}\fdens{\ftr} = \vzero, \quad\mbox{or}\quad
  \pvmufnc{\ftr}\fdens{\ftr} + 2\lambda \pvsmfnc{\ftr}\pportwfnc{\ftr}\fdens{\ftr} = \vzero.
\end{equation}
We need only describe the behavior of \pportwfnc{\ftr} for the case where $\fdens{\ftr} > 0$, as other $\ftr$ are
from a set of measure zero.
We follow common convention and state that a predicate holds ``almost surely'' if it holds for all \ftr except
possibly some \ftr where $\fdens{\ftr} = 0$.

Because the matrix $\pvsmfnc{\ftr}$ is a positive definite matrix, the first equation cannot hold unless
$\pportwfnc{\ftr} = \vzero$ almost surely. 
This would not satisfy the constraint, so we can ignore this possibility and can focus on the second equation, which reduces to 
\begin{equation}
\pportwfnc{\ftr} = c \pvinvsmfnc{\ftr}\pvmufnc{\ftr},
  \label{eqn:pporwtopt_soln_I}
\end{equation}
for some constant $c$, almost surely. 
The risk constraint of \problemref{opt_portsr_II} will fix the value of $c$ as 
\begin{equation}
\nonumber
  c = \frac{\Rbuj}{\sqrt{\qv - \qvpow{2}}},\quad\mbox{where}\quad
\qv \defeq \int \trpvmufnc{\ftr}\pvinvsmfnc{\ftr}\pvmufnc{\ftr}\fdens{\ftr}\dx[\ftr].
\end{equation}
As discussed laster, $\qvpow{1/2}$ is the (unconditional) \txtHR of the optimal portfolio allocation.


Thus we have
\begin{equation}
  \pportwoptfnc{\ftr} \defeq \frac{\Rbuj}{\sqrt{\qv - \qvpow{2}}} \pvinvsmfnc{\ftr}\pvmufnc{\ftr} \quad\mbox{a.s.}
  \label{eqn:opt_portsr_soln_I}
\end{equation}
We are assuming that $\pvinvsmfnc{\ftr}\pvmufnc{\ftr} \in \Ltwo$, which will require that \pvsmfnc{\ftr} have strictly positive eigenvalues, almost surely.
For this choice of portfolio function we have
\begin{align*}
  \pmufnc{\pportwopt} &= \Rbuj\sqrt{\frac{\qv}{1-\qv}},&
  \psmfnc{\pportwopt} &= \Rbuj^2\frac{1}{1-\qv},\\
  \pvarfnc{\pportwopt} &= \Rbuj^2,&
  \priskfnc{\pportwopt} &= \Rbuj.
\end{align*}
The objective takes optimal value
\begin{align}
  \label{eqn:optimal_objective}
  \frac{\pmufnc{\pportwopt}-\rfr}{\priskfnc{\pportwopt}} 
  &= \sqrt{\frac{\qv}{1-\qv}} - \frac{\rfr}{\Rbuj}.
\end{align}

\subsubsection{Hansen Ratio}

\citeauthor{ern2020hansen} defined the \txtHR as the mean return of a strategy divided by the square root of the second
moment of returns. \cite{ern2020hansen}
For our multi-period problem this is written as ${\pmufnc{\pportw}}/{\sqrt{\psmfnc{\pportw}}}$.
Repeating the diagram from above illustrates the connection between the \txtSR and the \txtHR:
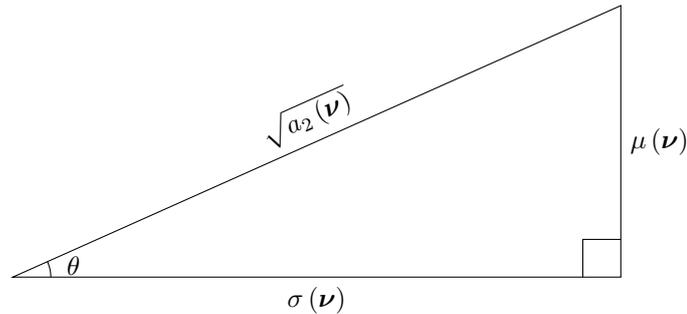
\begin{figure}[h]
  \centering
\begin{tikzpicture}
  \draw
  (1,0) coordinate (O) 
  -- (9,0) coordinate (A) 
  node[midway, below]{$\priskfnc{\pportw}$}
  -- (9,3.6) coordinate (B) 
  node[midway, right]{$\pmufnc{\pportw}$}
  -- cycle
  node[midway, above, sloped]{$\sqrt{\psmfnc{\pportw}}$}
  pic[draw, black] {right angle=O--A--B}
  (1.8,0.15) coordinate (T) node[]{$\theta$}
  pic[draw, black] {angle=A--O--B}
  ;
\end{tikzpicture}
\end{figure}

With $\theta$ the indicated angle of this triangle, the \txtSR of \pportw is $\tan{\theta}$ while its \txtHR is $\sin{\theta}$.
Note that this immediately establishes that the \txtHR never exceeds one in absolute value.
To maximize either of these quantities is equivalent to maximizing $\theta$.
Generally speaking when we seek a portfolio which maximizes the \txtSR, we can instead maximize the \txtHR.
Then in particular, \problemref{opt_portsr} has the same solution as 
\begin{equation}
\max_{\pportw\in\cone \,:\, \priskfnc{\pportw} \le \Rbuj} \frac{\pmufnc{\pportw}}{\sqrt{\psmfnc{\pportw}}}.
\label{problem:opt_portsmm}
\end{equation}
The optimal objective of this problem is \qvpow{1/2}.

If $\psnr$ is the \txtSR of a portfolio, and $h$ is its \txtHR then
\begin{equation}
\nonumber
  \psnr = \fntas{h} = \frac{h}{\sqrt{1-h^2}},\quad\mbox{and}\quad
  h = \fintas{\psnr} = \frac{\psnr}{\sqrt{1 + \psnrsq}}.
\end{equation}
Here ``tas'' stands for ``tangent of arcsin''.  \cite{pav_the_book}

Now rewrite 
the \txtSR of the optimal allocation from \eqnref{optimal_objective} as follows:
\begin{align}
  \frac{\pmufnc{\pportwopt}-\rfr}{\priskfnc{\pportwopt}} 
  &= \sqrt{\frac{\qv}{1-\qv}} - \frac{\rfr}{\Rbuj} =
\fntas{\qvpow{1/2}} - \frac{\rfr}{\Rbuj}.
\end{align}
This establishes that $\qvpow{1/2}$ is the \txtHR of the optimal portfolio, under zero risk-free rate.


We note that
\begin{align*}
\qv 
&= \int \trpvmufnc{\ftr}\pvinvsmfnc{\ftr}\pvmufnc{\ftr}\fdens{\ftr}\dx[\ftr],\\
  &= \Eof[\ftr]{\trpvmufnc{\ftr}\pvinvsmfnc{\ftr}\pvmufnc{\ftr}},\\
  &= \Eof[\ftr]{\frac{\trpvmufnc{\ftr}\pvinvsigfnc{\ftr}\pvmufnc{\ftr}}{1 + \trpvmufnc{\ftr}\pvinvsigfnc{\ftr}\pvmufnc{\ftr}}},\\
  &= \Eof[\ftr]{\frac{\psnrsqoptfnc{\ftr}}{1 + \psnrsqoptfnc{\ftr}}},\\
  &= \Eof[\ftr]{\qvfnc{\ftr}},
\end{align*}
where 
we define
$$\qvfnc{\ftr} \defeq \trpvmufnc{\ftr}\pvinvsmfnc{\ftr}\pvmufnc{\ftr} = \wrapParens{\fintas{\psnroptfnc{\ftr}}}^2.$$
Thus $q$, the squared (unconditional) \txtHR of the optimal portfolio is the expectation, over \ftr, 
of the squared conditional \txtHR of the optimal conditional portfolio.
Thus the (squared) \txtHR aggregates in a very natural way from the conditional to unconditional settings.
On the other hand, 
the aggregation of \txtSR (or squared \txtSR) from conditional to unconditional requires transformation by the ``tas'' function.





\subsection{Constrained Case}
\label{sec:constrained_case}

For this section we consider it more convenient to introduce the inner product notation.
For vector-valued functions $\vect{g}, \vect{h} \in \Ltwo$, define
\begin{equation}
\nonumber
  \inr{\vect{g}}{\vect{h}} = \int \funcit{\tr{\vect{g}}}{\ftr}\funcit{\vect{h}}{\ftr}\fdens{\ftr}\dx[\ftr].
\end{equation}

We now consider solutions to \problemref{opt_portsr_VI} where \cone is the intersection of a number of integral constraints:
\begin{equation}
\nonumber
  \cone = \setwo{\pportw}{\inr{\pportw}{\hejportw{j}} = 0,\, j=1,2,\ldots,J}.
\end{equation}
Here the $\hejportw{j}$ are given portfolio functions which we wish to be orthogonal, in expectation, to our selected portfolio.
Note that via this formulation we can capture things like a zero net dollar constraint 
(by taking $\hejportw{j}$ to be the function that is the constant ones vector).
We can also capture \emph{hedging constraints} where we wish to hold a portfolio whose returns have zero covariance
to some other portfolio, say $\vect{w}$. 
The zero covariance constraint is
\begin{align*}
  0 
  &= \inr{\pportw}{\pvsm\vect{w}} - \inr{\pportw}{\pvmu}\inr{\vect{w}}{\pvmu},\\
  &= \inr{\pportw}{\pvsm\vect{w} - \wrapParens{\inr{\vect{w}}{\pvmu}}\scalop \pvmu},
\end{align*}
which is of the requisite form.

Note that these constraints are \emph{in expectation}, and do not hold conditionally.
That is, the ``zero net dollar constraint'' does not constrain us to portfolio functions with $\trvone\pportwfnc{\ftr}=0$,
but instead the expectation, over \ftr, of $\trvone\pportwfnc{\ftr}$ is zero.
To get conditional constraints, one needs to further limit the space of acceptable allocation functions to
respect the constraints.

Thus in the constrained case we are seeking to solve the optimization problem
\begin{equation}
  \max_{\pportw\in\Ltwo} \inr{\pportw}{\pvmu} \quad\mbox{s.t.}\quad
  \inr{\pportw}{\pvsm\pportw} = 1,\,
  \inr{\pportw}{\hejportw{j}} = 0,\, j=1,2,\ldots,J.
\end{equation}
This is still an isoperimetric problem. 
The necessary conditions at a solution, as in the unconstrained case,
require that a linear combination of the integrands satisfy the Euler Lagrange equation.
That is
\begin{equation}
\nonumber
\gradof[\pportw]{N} = \vzero,\quad\mbox{where}\quad
\funcit{N}{\ftr, \pportw} = 
  \tr{\pportw}\pvmufnc{\ftr}\fdens{\ftr} + \lambda_0 \tr{\pportw}\pvsmfnc{\ftr}\pportw \fdens{\ftr} +
  \sum_j \lambda_j\tr{\pportw}{\hejportwfnc{j}{\ftr}}\fdens{\ftr}.
\end{equation}
Taking the gradient and 
factoring out the density of \ftr, the necessary condition is 
\begin{equation}
  \vzero = \pvmufnc{\ftr} + 2\lambda_0 \pvsmfnc{\ftr}\pportw + \sum_j \lambda_j \hejportwfnc{j}{\ftr}\quad\mbox{almost surely.}
\end{equation}
This has solution
\begin{equation}
\nonumber
\pportwoptfnc{\ftr} = c \pvinvsmfnc{\ftr} \wrapParens{\pvmufnc{\ftr} + \sum_j c_j \hejportwfnc{j}{\ftr}},
\end{equation}
where $c$ is some overall constant which will be set by the risk budget constraint.

For this solution to satisfy the hedging constraints we must further have
\begin{equation}
\nonumber
  0 = \inr{\hejportw{i}}{\pportwopt} = c \wrapParens{\inr{\hejportw{i}}{\pvinvsm\pvmu} + \sum_j c_j \inr{\hejportw{i}}{\pvinvsm\hejportw{j}}},
\end{equation}
for $i = 1, 2, \ldots, J$.
This imposes $J$ equality constraints for the $J$ unknowns $c_1, \ldots, c_J$,
since the $c$ factors out.
This can be expressed as a linear system:
\begin{equation}
\Mtx{M} \vect{c} = \vect{b},
\end{equation}
where 
\begin{equation}
  \MtxUL{M}{}{i,j} = \inr{\hejportw{i}}{\pvinvsm\hejportw{j}},\quad \vectUL{b}{}{i} = - \inr{\hejportw{i}}{\pvinvsm\pvmu}.
  \label{eqn:Mandb}
\end{equation}

Now we consider the unconditional first and second moments of this optimal portfolio. These are
\begin{align*}
  \inr{\pvmu}{\pportwopt} 
  &= c \wrapParens{\inr{\pvmu}{\pvinvsm\pvmu} + \sum_j c_j \inr{\pvmu}{\pvinvsm\hejportw{j}}},\\
  &= c \wrapParens{\inr{\pvmu}{\pvinvsm\pvmu} - \tr{\vect{b}}\minv{\Mtx{M}}\vect{b}},\\
  &= c \qv[g],\\
  \inr{\pportwopt}{\pvsm\pportwopt} 
  &= \inr{\pportwopt}{c \pvmu + c \sum_j c_j \hejportw{j}},\\
  &= c \inr{\pportwopt}{\pvmu} + c \sum_j c_j \inr{\pportwopt}{\hejportw{j}},\\
  &= c \inr{\pportwopt}{\pvmu},
\end{align*}
because $\pportwopt$ is orthogonal to all the $\hejportw{j}$, so the latter sum is over all zeros.
We define $\qv[g]$ to be the quantity in parentheses above; it corresponds to the optimal squared
\txtHR under the given constraints.
To saturate the risk budget we need
\begin{equation}
\nonumber
  \Rbuj^2 = c \inr{\pportwopt}{\pvmu} - {\inr{\pportwopt}{\pvmu}}^2 = c^2 \wrapParens{\qv[g] - \qvpow[g]{2}}.
\end{equation}
This fixes the identity of $c$ thus we have
\begin{equation}
\pportwopt = \frac{\Rbuj}{\sqrt{\qv[g] - \qvpow[g]{2}}} \pvinvsm \wrapParens{\pvmu + \sum_j c_j \hejportw{j}}.
\end{equation}
The \txtSR of this portfolio, including the risk-free rate, is
\begin{equation}
\nonumber
  \frac{\inr{\pportwopt}{\pvmu} - \rfr}{c \sqrt{\wrapParens{\qv[g] - \qvpow[g]{2}}}} 
  = \sqrt{\frac{\qv[g]}{1 - \qv[g]}} - \frac{\rfr}{\Rbuj}.
\end{equation}

Now consider the impact on the squared \txtHR from imposing the hedge constraints.
We have
\begin{align*}
\qv - \qv[g] &= \tr{\vect{b}}\minv{\Mtx{M}}\vect{b}.
\end{align*}
The right hand side can be interpreted as the ``loss'' in squared \txtHR incurred by imposing
the constraints.
We note that it can be interpreted as the squared \txtHR of a different constrained optimization problem:
suppose that one optimizes the \txtSR (or \txtHR) over portfolios of the form
\begin{equation}
\nonumber
\pportwfnc{\ftr} = \sum_j c_j \pvinvsmfnc{\ftr}\hejportwfnc{j}{\ftr}.
\end{equation}
Then the optimal squared \txtHR of this portfolio is $\tr{\vect{b}}\minv{\Mtx{M}}\vect{b}$, as we show in
\secref{linear_constrained_case}.
Thus there is a kind of Pythagorean theorem at work here, where the squared \txtHR of the unconditional
optimal portfolio is the sum of the squared \txtHRs of the two orthogonal optimal portfolios.

\subsubsection{Hedging Example}

Consider the simple case where there is a single constraint consisting of a portfolio which should be hedged out.
That is, one should have zero covariance against the portfolio \mktportwfnc{\ftr}.
As noted above this means
\begin{equation}
\nonumber
\hejportw{1} = {\pvsm\mktportw - \wrapParens{\inr{\mktportw}{\pvmu}}\scalop \pvmu}.
\end{equation}
To find the constant $c_1$ in \pportwopt we must solve the single equation
\begin{equation}
\nonumber
0 = {\inr{\hejportw{1}}{\pvinvsm\pvmu} + c_1 \inr{\hejportw{1}}{\pvinvsm\hejportw{1}}},
\end{equation}
or
\begin{align*}
  c_1 
  &= - \frac{\inr{\hejportw{1}}{\pvinvsm\pvmu}}{\inr{\hejportw{1}}{\pvinvsm\hejportw{1}}},\\
  &= - \frac{\inr{\mktportw}{\pvmu} - \inr{\mktportw}{\pvmu} \inr{\pvmu}{\pvinvsm\pvmu}}{
    \inr{\mktportw}{\pvsm\mktportw} - 2 \inr{\mktportw}{\pvmu}^2 + \inr{\mktportw}{\pvmu}^2 \inr{\pvmu}{\pvinvsm\pvmu}}.
\end{align*}
For this value of $c_1$ we have
\begin{align*}
  \qv[g] 
  &= \inr{\pvmu}{\pvinvsm\pvmu} + c_1 \inr{\pvmu}{\pvinvsm\hejportw{1}},\\
  &= \inr{\pvmu}{\pvinvsm\pvmu} - \frac{\inr{\pvmu}{\pvinvsm\hejportw{1}}^2}{\inr{\hejportw{1}}{\pvinvsm\hejportw{1}}},
\end{align*}
This can be further expanded in terms of the definition of \hejportw{1}, but it only makes the expression more complicated, rather
than less.

\subsection{Discrete Features}

Now consider the case where the $\ftr[t]$ is discrete, taking one of a finite number, \nstate, of values.
Without loss of generality it suffices to consider $\ftr[t]$ taking a value from $1$ to $\nstate$.
Let \pist[s] be the probability that we observe value $s$, which we will refer to as ``observing state $s$.''
Let \pvmu[s] be the conditional expected return of the assets conditional on observing state $s$,
and similarly let \pvsmFoo{s} be the conditional expected second moment, and
\pvsig[s] be the conditional covariance: $\pvsig[s] = \pvsmFoo{s} - \pvmu[s]\trpvmu[s]$.
Conditional on observing state $s$ we allocate into portfolio $\pportw[s]$.

The exposition above suggests that we should compute
\begin{equation}
  \qv = \sum_{1 \le s \le \nstate} \pist[s] \trpvmu[s]\pvinvsmFoo{s}\pvmu[s],
\label{eqn:discrete_q}
\end{equation}
then conditional on observing state $s$ to allocate to
\begin{equation}
\nonumber
  \pportwoptFoo{s,} = \Rbuj\sqrt{\frac{\qv}{1-\qv}} \pvinvsmFoo{s}\pvmu[s].
\end{equation}
Note that this is different than the conditional \txtMP, which would take value
$c {\pvinvsig[s]}\pvmu[s]$ for some constant $c$ independent of $s$.

As the proof based on calculus of variations has many steps, and may rely on unfamiliar machinery,
we show directly in this case that the optimal solution takes the form we wrote above.
The expected return and variance of returns of our portfolio is
\begin{align*}
  \pmufnc{\pportw} &= \sum_{1 \le s \le \nstate} \pist[s] \trpvmu[s]\pportw[s],\\
  \psmfnc{\pportw} &= \sum_{1 \le s \le \nstate} \pist[s] \trpportw[s]\pvsmFoo{s}\pportw[s].
\end{align*}
Starting from \problemref{opt_portsr_II}, the Lagrangian function is
\begin{equation}
\nonumber
  \lagrangfnc[\pportw, \lambda] = 
  \frac{\pmufnc{\pportw} - \rfr}{\priskfnc{\pportw}} + \lambda \wrapParens{\priskfnc{\pportw} - \Rbuj}.
\end{equation}
We now take the gradient with respect to the subvector $\pportw[s]$ and set to zero. This yields:
\begin{align*}
  \gradof[{\pportw[s]}]{\lagrangfnc[\pportw, \lambda]} 
  &= \frac{\pist[s]\pvmu[s]}{\priskfnc{\pportw}} 
  - \frac{\pmufnc{\pportw} - \rfr}{\pvarfnc{\pportw}}\gradof[{\pportw[s]}]{\priskfnc{\pportw}}
  + \lambda \gradof[{\pportw[s]}]{\priskfnc{\pportw}},\\
  &= c_1 \pist[s]\pvmu[s] + c_2 \gradof[{\pportw[s]}]{\priskfnc{\pportw}},\\
  &= c_1 \pist[s]\pvmu[s] + c_2 \wrapParens{\gradof[{\pportw[s]}]{\psmfnc{\pportw}} - 2\pmufnc{\pportw}\pist[s]\pvmu[s]},\\
  &= c_1 \pist[s]\pvmu[s] + c_2 \wrapParens{2\pist[s]\pvsmFoo{s}\pportw[s] - 2 \pmufnc{\pportw}\pist[s]\pvmu[s]},\\
  &= \pist[s] \wrapParens{c_1' \pvmu[s] + c_2' \pvsmFoo{s}\pportw[s]},
\end{align*}
where the constants $c_i'$ depend on $\pportw$ and $\lambda$, but do not depend on the state $s$.
Setting this to zero we can find $\pportwoptFoo{s,}$ up to scaling. 
Namely we have
\begin{equation}
\nonumber
\pportwoptFoo{s,} = c \pvinvsmFoo{s}\pvmu[s].
\end{equation}
The constant $c$ does not depend on $s$. It is a simple exercise to establish the identity of $c$ based on saturating the risk
budget.

We can rewrite \eqnref{discrete_q} via the Sherman-Morrison identity as
\begin{align}
  \qv 
\nonumber
  &= \sum_{1 \le s \le \nstate} \pist[s] \trpvmu[s]\pvinvsmFoo{s}\pvmu[s],\\
\nonumber
  &= \sum_{1 \le s \le \nstate} \pist[s] \trpvmu[s]\minvParens{\pvsig[s]+\pvmu[s]\trpvmu[s]}\pvmu[s],\\
\nonumber
  &= \sum_{1 \le s \le \nstate} \pist[s] \frac{\trpvmu[s]\pvinvsig[s]\pvmu[s]}{1 + \trpvmu[s]\pvinvsig[s]\pvmu[s]},\\
  &= 1 - \sum_{1 \le s \le \nstate} \pist[s] \frac{1}{1 + \trpvmu[s]\pvinvsig[s]\pvmu[s]}.
\label{eqn:discrete_q_II}
\end{align}
We would like to compare this to the \txtHR of the investor who holds the unconditional \txtMP in every period,
but found no obvious simplification of their difference or ratio. 
Instead we consider a simple example.


\subsubsection{An Example}

Consider the case of two assets, and $\nstate=2$ discrete states. 
Suppose that $\pist[1] = \pist[2] = 1/2$ and 
\begin{align*}
  \pvmu[1] &= \twobyone{1}{1},
  & \pvmu[2] &= \twobyone{2}{2},&
  \pvsig[1] &= \twobytwossym{1}{0}{1},
  &\pvsig[2] &= \twobytwossym{2}{0}{2}.
\end{align*}
Then we have
\begin{align*}
  \pportwoptFoo{1,} &= c\minv{\twobytwossym{2}{1}{2}}\twobyone{1}{1} = \twobyone{1/3}{1/3},
  & \pportwoptFoo{2,} &= c\minv{\twobytwossym{6}{4}{6}}\twobyone{2}{2} = \twobyone{1/5}{1/5}.
\end{align*}
Moreover we compute $\qv$ as
\begin{equation}
  \nonumber
  \qv = \frac{1}{2}\wrapParens{\frac{2}{3} + \frac{4}{5}} = \frac{11}{15}.
\end{equation}
Supposing $\Rbuj=1$ and $\rfr=0$, the objective of our optimization problem takes value
\begin{equation}
  \nonumber
  \sqrt{\frac{\qv}{1-\qv}} = \sqrt{\frac{11}{4}}.
\end{equation}

Now consider instead holding the conditional \txtMP in each of the two states.
That is, in both $s=1,2$ one holds $c\pvinvsig[s]\pvmu[s]$, or
\begin{equation*}
c \minv{\twobytwossym{1}{0}{1}}\twobyone{1}{1} = 
c \minv{\twobytwossym{2}{0}{2}}\twobyone{2}{2} = 
c \twobyone{1}{1}.
\end{equation*}
The mean return and risk of this strategy are 
\begin{align*}
\pmufnc{\pportw} &= \frac{c}{2}\wrapParens{2 + 4} = 3c,\\
\pvarfnc{\pportw} &= \frac{c^2}{2}\wrapParens{6 + 20} - 9c^2
  = c^2\wrapParens{13 - 9}
  = 4c^2,\\
\priskfnc{\pportw} &= 2c.
\end{align*}
Thus the objective of the conditional \txtMP is only $\frac{3}{2} < \frac{\sqrt{11}}{2}$.
The boost to the \txtSR by holding the \txtSMMP in this case is approximately 10.5\%.

\subsubsection{Omitted States}

The optimal objective value in the case of discrete features is 
\begin{equation}
\nonumber
\sqrt{\frac{\qv}{1-\qv}} - \frac{\rfr}{\Rbuj},
\end{equation}
where
\begin{equation}
\nonumber
\qv = \sum_s \pist[s] \trpvmu[s]\pvinvsmFoo{s}\pvmu[s].
\end{equation}
Now consider the case where one does not observe all \nstate states, 
rather whenever states $1$ or $2$ hold, one observes them as, say, state $3/2$.
We consider the impact on $\qv$.
The impact to $\qv$ will be
\begin{align*}
  \Delta \qv 
  &= \pist[3/2] \trpvmu[3/2]\pvinvsmFoo{3/2}\pvmu[3/2] - \sum_{s=1,2} \pist[s] \trpvmu[s]\pvinvsmFoo{s}\pvmu[s],\\
  &= \wrapParens{\pist[1] + \pist[2]} 
  \frac{\pist[1]\trpvmu[1] + \pist[2]\trpvmu[2]}{\pist[1]+\pist[2]} 
  \minvParens{\frac{\pist[1]\pvsmFoo{1} + \pist[2]\pvsmFoo{2}}{\pist[1]+\pist[2]}}
  \frac{\pist[1]\pvmu[1] + \pist[2]\pvmu[2]}{\pist[1]+\pist[2]} 
  - \sum_{s=1,2} \pist[s] \trpvmu[s]\pvinvsmFoo{s}\pvmu[s],\\
  &= \tr{\wrapParens{\pist[1]\pvmu[1] + \pist[2]\pvmu[2]}}
  \minvParens{\pist[1]\pvsmFoo{1} + \pist[2]\pvsmFoo{2}}
  \wrapParens{\pist[1]\pvmu[1] + \pist[2]\pvmu[2]}
  - \sum_{s=1,2} \pist[s] \trpvmu[s]\pvinvsmFoo{s}\pvmu[s],\\
  &= \pist[1]\wrapBracks{
    \tr{\wrapParens{\pvmu[1] + \gamma\pvmu[2]}}
  \minvParens{\pvsmFoo{1} + \gamma\pvsmFoo{2}}
  \wrapParens{\pvmu[1] + \gamma\pvmu[2]}
  - \trpvmu[1]\pvinvsmFoo{1}\pvmu[1]
  - \gamma \trpvmu[2]\pvinvsmFoo{2}\pvmu[2]},
\end{align*}
where $\gamma = \pist[2]/\pist[1]$.
The worst case largest drop in $\qv$ occurs when $\pvmu[1] = - \gamma\pvmu[2]$, and the contribution to $\qv$ of state $3/2$ is exactly zero. 
In that case we have
\begin{align*}
  \Delta q 
  &= \pist[1]\wrapBracks{
  - \trpvmu[1]\pvinvsmFoo{1}\pvmu[1]
  - \gamma \trpvmu[2]\pvinvsmFoo{2}\pvmu[2]},\\
  &= \pist[1]\wrapBracks{
  - \trpvmu[1]\pvinvsmFoo{1}\pvmu[1]
  - \frac{1}{\gamma} \trpvmu[1]\pvinvsmFoo{2}\pvmu[1]},\\
&= - \pist[1]\trpvmu[1]\wrapParens{\pvinvsmFoo{1} + \frac{1}{\gamma}\pvinvsmFoo{2}}\pvmu[1].
\end{align*}
This means that it is possible to have two different states such that if they were undifferentiated the impact on \qv would be
disastrous, with $\Delta \qv$ approaching $-1$.
Getting useful bounds for the omitted variable impact requires further assumptions.



\subsection{Optimizing Over Basis Portfolios}
\label{sec:linear_constrained_case}

Suppose that the function space of acceptable portfolio functions $\pportwfnc{\ftr}$ is spanned by a finite set of 
basis functions $\pportwifnc[i]{\ftr}$.
That is any portfolio function can be uniquely expressed as
\begin{equation}
  \nonumber
  \pportwfnc{\ftr} = \sum_{i} \beta_i \pportwifnc[i]{\ftr}.
\end{equation}
Then the portfolio optimization of \problemref{opt_portsr} can be expressed as finding the optimal values of the $\beta_i$.
First note that
\begin{equation}
  \nonumber
  \pmufnc{\sum_{i}\beta_i \pportwifnc[i]{\ftr}} 
  = \sum_{i}\beta_i \pmufnc{\pportwifnc[i]{\ftr}}
\end{equation}
by linearity of the integral.
Moreover, 
\begin{align*}
  \psmfnc{\sum_{i} \beta_i \pportwifnc[i]{\ftr}} 
  &= \int \tr{\wrapParens{\sum_i \beta_i \pportwifnc[i]{\ftr}}} \pvsmfnc{\ftr} \wrapParens{\sum_j \beta_j \pportwifnc[j]{\ftr}} \fdens{\ftr}\dx[\ftr],\\
  &= \sum_{i,j} \beta_i \beta_j \int \tr{\pportwifnc[i]{\ftr}} \pvsmfnc{\ftr} \pportwifnc[j]{\ftr} \fdens{\ftr}\dx[\ftr],
\end{align*}
thus the second moment is bilinear in the $\beta$.
We can write this mean and second moment as operations on the vector $\bvec$ of $\beta_i$ coefficients as
\begin{align*}
  \pmufnc{\sum_i \beta_i \pportwifnc[i]{\ftr}} &= \tr{\bvec}\pvmu,\\
  \psmfnc{\sum_i \beta_i \pportwifnc[i]{\ftr}} &= \tr{\bvec}\pvsm\bvec,
\end{align*}
where we define the vector \pvmu and matrix \pvsm via
\begin{align*}
  \pmu[i] &= \int \tr{\pvmufnc{\ftr}} \pportwifnc[i]{\ftr}\fdens{\ftr}\dx[\ftr],\\
  \pvsmFoo{i,j} &= \int \tr{\pportwifnc[i]{\ftr}} \pvsmfnc{\ftr} \pportwifnc[j]{\ftr} \fdens{\ftr}\dx[\ftr].
\end{align*}

Let $\pvsig = \pvsm - \pvmu\trpvmu$.
Then \problemref{opt_portsr} can be expressed as 
\begin{equation}
  \max_{\bvec: \tr{\bvec}\pvsig\bvec \le \Rbuj^2} \frac{\tr{\bvec}\pvmu - \rfr}{\sqrt{\tr{\bvec}\pvsig\bvec}}.
  \label{problem:opt_portsr_linear_I}
\end{equation}
This looks like a typical portfolio optimization problem with a finite number of assets,
which in this case are the portfolios $\pportwifnc[i]{\ftr}$.

Introducing the Lagrange multiplier, and after some simplification, the necessary condition of a solution to this problem is that
\begin{equation*}
  c_1 \pvmu + c_2 \pvsig \bvec = 0,
\end{equation*}
and thus the problem is solved by $\bvec = c \pvinvsig\pvmu$.
To saturate the risk budget we take 
$$
c = \frac{\Rbuj}{\psnropt},
$$
where $\psnrsqopt=\trpvmu\pvinvsig\pvmu$.
Using the Sherman-Morrison identity as above this can be equivalently written as 
$$
\bvec = \frac{\Rbuj\wrapParens{1 + \psnrsqopt}}{\psnropt} \pvsm^{-1}\pvmu.
$$
The maximized objective of this portfolio is $\psnropt - \frac{\rfr}{\Rbuj}$.
The squared \txtHR of this optimal portfolio is
$$
\tr{\pvmu}\pvinvsm\pvmu.
$$

Now suppose the basis functions are all of the form
\begin{equation}
\pportwifnc[i]{\ftr} = 
  \pvinvsmfnc{\ftr}\hejportwfnc{j}{\ftr}.
\end{equation}
Then the vector \pvmu is equal to $-\vect{b}$ as given in \eqnref{Mandb},
and the matrix \pvsm is equal to the matrix \Mtx{M} in \eqnref{Mandb}.
This confirms the Pythagorean theorem quoted in the section on hedging:
the optimal squared \txtHR is the sum of the squared \txtHR on the spanned space
and the squared \txtHR over the orthogonal space.

\subsubsection{Linear Portfolio Functions}

Consider the case where the target portfolios are to be linear in the features $\ftr[t]$.
That is $\pportwfnc{\ftr[t]} = \ppasthru \ftr[t]$ for an appropriately sized matrix $\ppasthru$.
This is just a special case of the analysis above where the basis portfolio functions are
$\pportwifnc[i,j]{\ftr[t]} = \vone[i]\trvone[j]\ftr[t]$.
Then we have 
\begin{align}
  \pmufnc{\pportwifnc[i,j]{\ftr}} &= \int \trpvmufnc{\ftr}\vone[i]\trvone[j]\ftr\fdens{\ftr}\dx[\ftr],
\end{align}
which takes the expected value of the  conditional expected value of the \kth{i} asset with respect to the \kth{j} element of the
features.
That is, $\pmufnc{\pportwifnc[i,j]{\ftr}}$ is the unconditional expected value of the \kth{i,j} element of 
the ``pseudo-asset'', $\vreti[t]\kron\ftr[t]$.
Similarly the second moment matrix is the second moment matrix on the pseudo-assets.
Thus one can simply perform classical portfolio optimization techniques on the pseudo-assets $\vreti[t]\kron\ftr[t]$,
a method known as the ``flattening trick'' or ``augmenting the asset space''.
\cite{pav_the_book,JOFI:JOFI1055}

\subsection{Mean Variance Optimization}

Consider now the mean variance optimization problem
\begin{equation}
  \max_{\pportw\in\Ltwo} \pmufnc{\pportw} - \half[\riskt]\pvarfnc{\pportw}.
\label{problem:opt_portmvo}
\end{equation}
Suppose \pportwoptfnc{\ftr} is some function which solves \problemref{opt_portmvo},
and let $\Rbuj = \priskfnc{\pportwopt}$.
Then we claim that $\pportwopt$ is a solution to 
\begin{equation}
\max_{\pportw\in\Ltwo \,:\, \priskfnc{\pportw} = \Rbuj} \pmufnc{\pportw} - \half[\riskt\Rbuj^2].
\label{problem:opt_portmvo_III}
\end{equation}
Now note that \problemref{opt_portmvo_III} is 
equivalent to \problemref{opt_portsr_IV},
which has solution of the form given in \eqnref{pporwtopt_soln_I},
except that the constant $c$ may be computed differently.
Thus, as in the case of Sharpe maximization, the optimal portfolio is the Sherman-Morrison Markowitz portfolio.

Consider now the identity of the scaling constant $c$. 
We wish to solve \problemref{opt_portmvo} subject to $\pportwfnc{\ftr} = c \pvinvsmfnc{\ftr}\pvmufnc{\ftr}$, which is to say
\begin{equation}
  \max_{c} \quad c \pmufnc{\pportwopt} - \frac{c^2}{\riskt}\pvarfnc{\pportwopt},
\label{problem:opt_portmvo_scalar}
\end{equation}
where $\pportwoptfnc{\ftr} = \pvinvsmfnc{\ftr}\pvmufnc{\ftr}$ is the unscaled \txtSMMP. 
By basic calculus this is solved when
\begin{equation}
c = \frac{\riskt\pmufnc{\pportwopt}}{2\pvarfnc{\pportwopt}} = \frac{\riskt\qv}{2\wrapParens{\qv - \qvpow{2}}} = \frac{\riskt}{2\wrapParens{1-\qv}}.
\end{equation}
The optimal value of the objective for this portfolio is equal to
\begin{equation}
\nonumber
\frac{\riskt}{4}\frac{\qv}{1 - \qv}.
\end{equation}

\subsubsection{Kelly Criterion}

The Kelly Criterion is a bet-sizing scheme devised in the 1950's 
designed to maximize expected log terminal wealth.  \cite{kelly1956,rotando_kelly_1992,doi:10.1142/7598}
We will employ the usual quadratic expansion of the logarithm to show
that a Kelly investor should also hold the \txtSMMP, if they can ignore skewness
and higher order moments.

Let $y_t$ be the simple returns gained by an investor after time period $t$.
The Kelly criterion is based on maximizing
\begin{equation}
  \nonumber
  \sum_t \Eof{\flog{1 + y_t}}.
\end{equation}
The expectation is over the realizations of $y_t$.
In our formulation, the investor observes $\ftr[t]$ prior to the investment decision,
in response to which they allocate $\pportwfnc{\ftr[t]}$.
Their returns are then $y_t = \trvreti[t]\pportwfnc{\ftr[t]}.$

First, by the ``tower rule'' of expectations, 
\begin{align*}
  \Eof{\flog{1 + y}} 
  &= \Eof[\ftr]{\Econd{\flog{1+\trvreti\pportwfnc{\ftr}}}{\ftr}}.
\end{align*}
Now we use the quadratic expansion of the log, namely $\flog{1 + \epsilon} \approx \epsilon - \half \epsilon^2$
to rewrite the expectation:
\begin{align*}
  \Eof{\flog{1 + y}} 
  &\approx \Eof[\ftr]{\Econd{\trvreti\pportwfnc{\ftr} - \half \wrapParens{\trvreti\pportwfnc{\ftr}}^2}{\ftr}},\\
  &= \Eof[\ftr]{\trpvmufnc{\ftr}\pportwfnc{\ftr} - \half \trpportwfnc{\ftr}\pvsmfnc{\ftr}\pportwfnc{\ftr}},\\
  &= \pmufnc{\pportw} - \half \psmfnc{\pportw}.
\end{align*}
And thus a Kelly investor who accepts this approximation will seek to solve the portfolio problem
\begin{equation}
\max_{\pportw\in\Ltwo} \pmufnc{\pportw} - \half \psmfnc{\pportw},
  \label{problem:opt_kelly}
\end{equation}
As with the mean variance analysis above, if we let $r = \psmfnc{\pportwopt}$ for the
optimal portfolio function $\pportwoptfnc{\ftr}$, then this problem is equivalent to 
a variant of \problemref{opt_portsr_VI}.
Once again the solution is to hold the \txtSMMP as given in \eqnref{pporwtopt_soln_I}.
Again, the leading constant $c$ can be found, via simple calculus, to be
$$
c = \frac{\pmufnc{\pportwopt}}{\psmfnc{\pportwopt}} = \frac{\qv}{\qv} = 1.
$$
And thus the (approximate) Kelly investor will hold the unscaled \txtSMMP, $\pvinvsmfnc{\ftr}\pvmufnc{\ftr}$.
The objective value of this portfolio is $\qv/2$.
A ``fractional Kelly'' investor will hold some down-levered fraction of the full \txtSMMP,
to reduce the probability of a single disastrous loss.
 \cite{doi:10.1142/7598}

\section{Applications}

We briefly mention a few practical applications of the theory of the \txtSMMP.

\subsection{Linear Conditional Expectation Model}

Consider the \emph{linear conditional expectation model}, \cite{pav2013markowitz}
\begin{align*}
  \pvmufnc{\ftr[t]} &= \pRegco\ftr[t],\\
  \pvsmfnc{\ftr[t]} &= \pvsig + \ogram{\wrapParens{\pRegco\ftr[t]}}.
\end{align*}
The optimal portfolio, by \eqnref{opt_portsr_soln_I} is
\begin{align*}
  \pportwoptfnc{\ftr[t]} 
&= \frac{\Rbuj}{\sqrt{\qv - \qvpow{2}}} \frac{1}{1 + \qform{\pvinvsig}{\wrapParens{\pRegco\ftr[t]}}} \pvinvsig\pRegco\ftr[t].
\end{align*}
The optimal objective for this portfolio is
\begin{equation}
\nonumber
\sqrt{\frac{\qv}{1-\qv}} - \frac{\rfr}{\Rbuj},
\end{equation}
where $\qv$ takes value
\begin{align*}
\qv 
&= \int \qiform{\wrapParens{\pvsig + \ogram{\wrapParens{\pRegco\ftr}}}}{\wrapParens{\pRegco\ftr}} \fdens{\ftr}\dx[\ftr],\\
&= \int \frac{\qiform{\pvsig}{\wrapParens{\pRegco\ftr}}}{1 + \qiform{\pvsig}{\wrapParens{\pRegco\ftr}}}\fdens{\ftr}\dx[\ftr].
\end{align*}

In contrast the conditional \txtMP is equal to 
\begin{equation}
\nonumber
c \pvinvsig\pRegco\ftr[t]
\end{equation}
for some constant $c$. The unconditional mean and variance of this allocation are
\begin{align*}
  \pmu &=c \int \tr{\wrapParens{\pRegco\ftr}}\pvinvsig\pRegco\ftr \fdens{\ftr}\dx[\ftr],\\
  &=c \trace{\qform{\pvinvsig}{\pRegco} \Eof{\ogram{\ftr}}}.\\
  \psigsq &=\int \qform{\wrapParens{\pvsig + \ogram{\wrapParens{\pRegco\ftr}}}}{\wrapParens{c\pvinvsig\pRegco\ftr}} \fdens{\ftr}\dx[\ftr] -
  \pmu^2,\\
  &= c\pmu 
  + c^2 \Eof{\wrapParens{\trace{\qform{\pvinvsig}{\pRegco} \ogram{\ftr}}}^2} 
  - \pmu^2.
\end{align*}

\subsubsection{An Example}

Further simplificiation of the equations above to compare the performance gap between the \txtSMMP and the \txtMP does not seem easily attained.
However, we can compare them numerically based on some population data that we concocted.
Suppose that
\begin{align*}
  \ftr[t] &\sim \normlaw{\threebyone{1}{1}{-2},\eye[3]},\\
  \pRegco &= \twobythree{0.04}{0.02}{-0.03}{-0.03}{-0.02}{0.02},\\
  \pvsig &= \twobytwo{1}{-0.1}{-0.1}{1}.
\end{align*}
Via Monte Carlo simulations we estimate the integrals and find that the unconditional \txtSR of the \txtSMMP is
$$
\psnropt = \sqrt{\frac{\qv}{1 - \qv}} \approx 0.156.
$$
Meanwhile the manager who holds the conditional \txtMP in every period enjoys nearly the same \txtSR.
The difference in \txtSRs is estimated to be only
$$
\Delta \psnropt \approx \ensuremath{3.724\times 10^{-5}},
$$
a value so small it will have no practical effect.
This result will come as no surprise when we consider the effect of rescaling constant, 
$$
\minvParens{1 + \qiform{\pvsig}{\wrapParens{\pRegco\ftr}}}.
$$
We estimate the standard deviation of this rescaling constant in this example to be only $0.018$,
thus there is little average difference between the scale of the conditional \txtMP and that of the \txtSMMP.


\subsection{A Neural Net Recipe}

One failing of the linear conditional expectation model is that it does not recognize the predictable changes in volatility
which are often visible in asset returns. \cite{stylized_facts,ARCH1987}
Another weakness is that it forces us to perform featurizations of observable data or assume that linear functions are good enough.
Both of these defects can be addressed by using a neural net to approximate the functions \pvmufnc{\ftr} and \pvsmfnc{\ftr}.
We provide a high level recipe for doing so, recognizing that there are myriad omitted details.
While mathematically the features \ftr[t] are expressed as a vector, in reality these variables are likely to come in two
forms: those which are specific to the assets, and those which are ``macroeconomic'' or otherwise apply to all assets.
For example, when considering equities trading we expect features to consist of a bunch of time-by-stock matrices,
perhaps measured at different frequencies; the latter can be expressed as a collection of single time series.
The neural net should be designed to ingest these, perhaps keeping applying the same computations to the per-stock features,
and otherwise ingesting the macroeconomic features.
The output should consist of ``heads'' for \pvmufnc{\ftr} and \pvsmfnc{\ftr}.
Likely the latter should be computed as some low rank update to a diagonal matrix.
Because of the danger of overfitting we recommend the network contain some low dimensional ``bottleneck'' between features
and output.
To perform model fitting, one could assume some elliptical distribution of asset returns then maximize likelihood.

\subsection{Investigating Leverage}

One odd possible application of the theory is in investigating whether an existing strategy makes optimal use of leverage.
Alternatively one can view this as creating a kind of overlay which acts on top of an existing strategy.
It works as follows: suppose you observe the period returns of a strategy, call them $z_t$, as well as the leverage of the
strategy, defined as the sum of absolute proportional allocation in each asset. Denote this leverage by $x_t$.
Now consider the returns of the unit-levered version of the strategy, defined as
$$
y_t = \frac{z_t}{x_t}.
$$
While the feature and returns are scalars instead of vectors, we can think of this problem in the same framework as above.
Since the strategy somehow chooses to allocate to total leverage of $x_t$, this feature is clearly observable to us
prior to the investment decision. The optimal leverage in each period is 
$$
c \frac{\pmufnc{x_t}}{\psmfnc{x_t}}.
$$
If we could estimate this function, then plot it against $x_t$, we should hope to see a straight line through the origin.
We can estimate the numerator and denominator in this fraction separately,
perhaps via non-parametric techniques, since they correspond to how $y_t$ and $y_t^2$ vary with $x_t$ in a given sample.
Care should be taken to force the denominator to be non-negative.

\section{Conclusion and Future Directions}

We established optimality of the \txtSMMP for portfolio optimization problems under the unconditional \txtSR objective as well as
the mean-variance objective, including approximate Kelly criterion optimization.
The \txtSMMP differs from the conditional \txtMP in each period by relatively down-levering when the conditional squared
\txtSR is higher.
We show that the optimal squared \txtHR is the expected value of the conditional squared \txtHR; 
in the multi-period context we find that replacing the centered variance or covariance matrix
with the uncentered versions is more natural and simplifies certain computations.
We separately prove the result in the discrete feature case, confirming the result.
We show how to deal with constraints in expectation, as well as how to optimize over a finite set of basis portfolios.
We establish a Pythagorean theorem for squared \txtHR of the spanned and orthogonal optimal portfolios.
Simulations show that the \txtSMMP only provides modest improvements over the \txtMP in terms of achieved \txtSR.
The practical impacts of switching to the \txtSMMP will likely be small, but one may find some solace in holding the optimal
portfolio.

We envision the following for further revisions of this work:
\begin{compactenum}
\item This work assumes that the functions $\pvmufnc{\ftr[t]}$ and $\pvsmfnc{\ftr[t]}$ are known, while in reality they have to be
  estimated from data.
  While estimation of the mean is fairly straightforward under most commonly assumed models, modeling the second moment matrix
  function is unusual. 
  Further work would establish the correct way to do this.
\item Similarly, while much is known about performing inference on the \txtSR \cite{pav_the_book}, much less is known about doing
  so on the \txtHR, either in the conditional case, or in the unconditional optimal allocation we outline here.
  Moreover, if one could trade on one of two different universes of assets, with different observational histories, how should one
  make that decision? How does estimation error affect the achieved outcomes?
\item We suspect one can establish bounds on the gap in performance between the \txtMP and the \txtSMMP. 
  Doing so would help practitioners judge the value of switching to the \txtSMMP.
\item Our neural net recipe lacks specifics; more battle-tested recommendations would be welcome.
\item Similarly, it would be interesting to concoct a traditional ``linear'' model that takes into account conditional
  heteroskedasticity and compare it against a neural net recipe. 
\end{compactenum}


\bibliographystyle{plainnat}
\bibliography{smm}

\end{document}